# Energy-Momentum of the Gravitational Field:

# Crucial Point for Gravitation Physics and Cosmology

© **Yu. Baryshev**[1,2]

[1] Astronomical Institute of the St.-Petersburg State University, St.-Petersburg, Russia
[2] Email: yuba@astro.spbu.ru

**Abstract:** A history of the problem of mathematical and physical definition for the energy-momentum of the gravity field is reviewed. As it was noted 90 years ago by Hilbert (1917), Einstein (1918), Schrodinger (1918) and Bauer (1918) within Geometrical Gravity approach (General Relativity) there is no tensor characteristics of the energy-momentum for the gravity field. Landau & Lifshitz (1971) called this quantity pseudo-tensor of energy-momentum and noted that Einstein's equations does not express the energy conservation for matter plus gravity field. This has crucial consequences for gravity physics and cosmology, such as negative energy density for static gravity field and violation of energy conservation in expanding space. However there is alternative Field Gravity approach for description of gravitation as a symmetric tensor field in Minkowski space, which is similar to description of all other physical interactions and based on well-defined positive, localizable energy-momentum of the gravity field. This relativistic quantum Field Gravity approach was partially developed by Firz & Pauli (1939), Birkhoff (1944), Thirring (1961), Kalman (1961), Feynman (1963) and others. Here it is shown that existence of well-defined positive energy-momentum of the gravity field leads to radical changes in gravity physics and cosmology, including such new possibilities as two-component nature of gravity – attraction (spin 2) and repulsion (spin 0), absence of black holes and singularities, scalar gravitational radiation caused by spherically symmetric gravitational collapse.

## 1. The gravity energy-momentum pseudo-tensor in general relativity

*Origin of the problem.*

The geometrical way to gravity physics was built by Einstein (1915, 1916) in his general relativity (GR). Wheeler termed this approach geometrodynamics, underlining the fact that geometry is not a passive background but becomes a dynamical physical entity that may be deformed, stretched and even spread in the form of gravitational waves. Geometrical gravity is related to the curvature of space-time itself while other physical interactions are related to the matter in the flat Minkowski space-time.

The problem of the energy of the gravity field in general relativity has a long history, it was, in fact, born together with Einstein's equations. Hilbert (1917) was the first who noted that "I contended ... in general relativity ... no equations of energy ... corresponding to those in orthogonally invariant theories". Here "orthogonal invariance" refers to theories in the flat Minkowski space.

Emmy Noether (1918), a pupil of Hilbert, proved that the symmetry of Minkowski space is the cause of the conservation of the energy-momentum tensor of a physical field. Many results of modern relativistic quantum field theories are based on this theorem. So the "prior geometry" of the Minkowski space in the field theories has the advantage that it guarantees the tensor character of the energy-momentum and its conservation for the fields.

In fact, Einstein & Grossmann (1913) came close to Noether's result when they wrote: "remarkably the conservation laws allow one to give a physical definition of the straight line, though in our theory there is no object or process modeling the straight line, like a light beam in ordinary relativity theory". In other words, they stated that the existence of conservation laws implies the flat Minkowski geometry. In the same article Einstein & Grossmann also emphasized that the gravity field must have an energy-momentum tensor as all other physical fields. However, in the final version of general relativity Einstein rejected this requirement in order to have a generally covariant gravity theory with no prior Minkowski geometry.

Schrodinger (1918) showed that the mathematical object $t^{ik}$ suggested by Einstein in his final general relativity for describing the energy-momentum of the gravity field may be made vanish by a coordinate transformation for the Schwarzschild solution if that solution is transformed to Cartesian coordinates. Bauer (1918) pointed out that Einstein's energy-momentum object, when calculated for a flat space-time but in a curvilinear system of coordinates, leads to a nonzero result. In other words, $t^{ik}$ can be zero when it should not be, and can be nonzero when it should.

Einstein (1918) replied that already Nordstrom informed him about this problem with $t^{ik}$. Einstein noted that in his theory $t^{ik}$ is not a tensor and also it is not symmetric. He also withdrew his previous demand of the necessity to have an energy-momentum tensor: "There may very well be gravitational fields without stress and energy density".

The "pseudo-tensor" character of the gravity field in GR has been discussed from time to time for a century, causing surprises for each new generation of physicists. Rejecting the Minkowski space inevitably leads to deep difficulties with the definition and conservation of the energy-momentum for the gravity field.

*Mathematical formulation of the problem.*

According to general relativity (GR) gravity is described by a metric tensor $g^{ik}$ of a Riemannian space. The "field" equations in GR (Einstein-Hilbert equations) have the form (we use Landau & Lifshitz 1971 notations):

$$R^{ik} - \frac{1}{2} g^{ik} R = \frac{8\pi G}{c^4} T_m^{ik} , \qquad (1)$$

where $R^{ik}$ is the Ricci tensor, $R$ is the scalar curvature, $T_m^{ik}$ is the energy-momentum tensor (hereafter EMT) for all kind of matter and fields, including EMT of the dark energy and cosmological vacuum. The most important feature of the Einstein's equations is that the right part of eq.(1) does not include the energy-momentum of the gravity field itself, and this corresponds the fact that in GR the gravity is not a material field.

Due to the Bianchi identity, a direct mathematical consequence of Einstein's eq.(1) is that the covariant divergence of the EMT of the matter equals zero, i.e.

$$(T_m^{ik})_{;k} = \frac{1}{\sqrt{-g}} \frac{\partial(\sqrt{-g}\, T_m^{ik})}{\partial x^k} - \frac{1}{2} \frac{\partial g_{kl}}{\partial x_i} T_m^{kl} = 0 , \qquad (2)$$

One is tempted to see in this expression a usual conservation law, but let us cite the famous, but often ignored statement by Landau & Lifshitz (1971, sect.101): "however, this equation does not generally express any conservation law whatever. This is related to the fact that in a gravitational field the four-momentum of the matter alone must not be conserved, but rather the four-momentum of matter plus gravitational field; the latter is not included in the expression for $T_m^{ik}$ ".

Mathematically this is because the integral $\int T^{ik} \sqrt{-g}\, dS_k$ is conserved only if the condition $\partial(\sqrt{-g}\, T^{ik})/\partial x^k = 0$ is fulfilled, however the second term in eq.(2) generally does not allow it. To define a conserved total four-momentum for a gravitational field plus the matter within it, Landau & Lifshitz suggested the expression

$$\frac{\partial}{\partial x^k}(-g)(T_m^{ik} + t_g^{ik}) = 0 , \qquad (3)$$

here $t_g^{ik}$ is called the energy-momentum pseudo-tensor, which should describe energy density of the gravity field itself. It is important that the quantities $t_g^{ik}$ do not constitute a tensor, i.e. they depend on the choice of the system of coordinates, and this fact leads to mentioned above confusions. There are many suggestions for the pseudo-tensor, among them Einstein's (non-symmetric) and Landau & Lifshitz's (symmetric) pseudo-tensors.

However, this way of introducing energy for the gravity field is physically inconsistent, as discussed in detail by Logunov & Folomeshkin (1977) and Logunov & Mestvirishvili (1989). Moreover Yilmaz (1992) has shown that due to the Freud identity for the pseudo-tensor: $\partial_i(\sqrt{-g}\, t_k^i) = 0$ , there is a difficulty with the definition of the gravitational acceleration.

*Non-localizability of gravity energy.* There are attempts to give a physical reason for the non-tensor character of the energy of a gravity field, e.g. due to the non-localizability of the gravity field in the geometrical approach (Misner, Thorne & Wheeler 1973, p.467): "It is not localizable. The equivalence principle forbids." They also noted the following properties of the pseudo-tensor: "There is no unique formula for it, ... , 'local gravitational energy momentum' has no weight. It does not curve space. It does not serve as a source term ... It does not produce any relative geodesic deviation of two nearby world lines ... It is not observable." So the actual cause of the absence of the gravity energy, i.e. the pseudotensor character of the EMT of the gravitational field in general relativity, is the principle of equivalence, i.e. the geometrization principle.

However the problem remains, why all other fields are localizable, i.e. detectable by means of a local transformation of the field energy into the energy of a test particle. How can one detect, localize, hence extract energy from a non-localizable field by means of an antenna, like in gravitational wave detectors? If there is no local energy density of the field, then there is no energy in a finite volume, too. Absence of energy density of the gravity field in GR leads also to the problem of quantization of the gravity field, i.e. the energy of gravitons, quanta of the field. In Friedmann cosmology the problem of the energy of the gravity field leads to the paradox of continuous creation (annihilation) of matter within any finite comoving volume.

## 2. Attempts to solve the energy-momentum problem within geometrical approaches

In the literature one may find several attempts to consider a combination of general relativity with a field approach by accepting some Lorentz-covariant properties of Minkowski space in "effective" Riemannian space (this is comprehensively reviewed by Pitts & Schieve 2001). As an example of such works we mention three "field gravity theories" developed by Logunov, Yilmaz, Grishchuk, and their collaborators.

Logunov & Mestvirishvili (1989) developed field gravity theory, called the relativistic theory of gravitation (RTG), where besides the metric tensor $g^{ik}$ of the effective Riemann space, they introduced a "causality principle" as an additional restriction on $g^{ik}$. Because of this there is no black hole solution in RTG. There is also the scalar part of gravitational tensor potentials, but it exists only in a static field and can not be radiated. The cosmological solution is the Friedmann expanding space with the critical matter density.

Yilmaz (1992) constructed a field theory where the right-hand side of the field equation contains implicitly the EMT of the gravity field. The metric of the effective Riemann space has an exponential form and excludes the event horizon and singularity. The existence of the EMT of the gravity field allows one to consider N-body solutions in this theory.

Baback & Grishchuk (2000) claimed that the field approach and general relativity are completely identical: "GR may be formulated as a strict non-linear field theory in flat space-time. This is a different formulation of the theory, not a different theory." They introduce the metric tensor $g^{ik}$ of a curved space-time in the form $g^{ik} = (\eta^{ik} + h^{ik})\sqrt{\gamma/g}$, which is the sum of two non-tensor quantities – the Minkowski metric $\eta^{ik}$ and the field variable $h^{ik}$. Then they developed a Lagrangian theory which contains a tensor quantity identified with an energy-momentum tensor of the gravity field variable (close to LL-pseudotensor). Their theory contains also black holes and expanding space cosmology.

The internal inconsistency of such an approach was noted by Straumann (2000) who emphasized that in GR there is a non-trivial topology of space-time of a black hole, while Minkowski space has a trivial topology of the flat space-time. Also the expanding space violates energy conservation, which is impossible for the field in Minkowski space.

Common disadvantage of above approaches is also the lack of required physical properties of the gravity field EMT. Indeed, from the quantum relativistic field theories of other physical interactions it is known that the EMT of a massless boson field obeys the following conditions:
- *symmetry*, $T^{ik} = T^{ki}$;
- *positive energy density for static and free field*, $T^{00} > 0$;
- *zero trace*, $T = 0$.

The above considered attempts to introduce the EMT of the gravity field within geometrical and effective "field" approachs, though obey the symmetry condition, do not possess the other two necessary features of the EMT, i.e. a positive energy density and tracelessness. These must be fulfilled within the consistent field approach for both static and free fields, as is the case of the electromagnetic field.

A violation of the positiveness of the energy density of the gravity field within the above theories may be demonstrated with the simplest case of a spherically symmetric weak static gravity field. Indeed, for this case, like in a terrestrial laboratory, one can easily calculate the predicted value of the energy density of the gravitational field for different pseudo-tensors and suggested tensors. For instance, in harmonic coordinates the Landau-Lifshiz symmetric pseudotensor and the energy-momentum tensor of the gravity field, which was found by Grishchuk, Petrov & Popova (1984), has a negative energy density of the weak static field :

$$t_{LL}^{00} = -\frac{7}{8\pi G}\left(\vec{\nabla}\phi_N\right)^2, \quad \text{and} \quad t_{GPP}^{00} = -\frac{11}{8\pi G}\left(\vec{\nabla}\phi_N\right)^2 . \quad (4)$$

The negative energy density of the static gravitational field is in conflict with the quantum field theories of other fundamental interactions. Also the traces of all these EMTs do not vanish for static fields.

The above discussion demonstrates that all theories which introduce an effective metric of Riemannian space $g^{ik} \approx \eta^{ik} + h^{ik}$, (plus geometrical condition $g_k^i = \delta_k^i$ ), lose some essential properties of the consistent field approach (e.g. the scalar part of the gravitational potential and the necessary properties of the EMT) and receive some unphysical properties of the geometrical approach (e.g. the negative energy of the field, the event horizon and the expending space). Actually such effective field theories introduce different restrictions on the metric of Riemannian space (e.g. null cone conditions) or extend geometrical Einstein's equations.

As discussed by Padmanabhan (2004), all attempts to derive "geometry" from "field" explicitly or implicitly contain propositions that reduce the field approach to geometry. Hence, these physical theories are non-equivalent. The main question is still there --- how to construct a consistent field gravity theory, quantum gravidynamics, based on relativistic quantum principles and which only as an approximation to reality contains geometrical interpretation, like geometrical optics in quantum electrodynamics.

### 3. Field gravity approach based on positive energy density of the gravity field

*Poincare--Feynman field approach.*

There is another alternative way to construct a consistent physical gravity theory, which utilizes, similarly to other fundamental physical fields, Minkowski space and the Lagrangian formalism of the relativistic quantum field theory.

As early as 1905 Poincare in his work "On the dynamics of the electron" put forward an idea about a relativistic theory for all physical interactions, including gravity, in flat 4-d space-time (now called Minkowski space). He pointed out that analogously to electrodynamics, gravitation should propagate with the velocity of light, and there should exist mediators of the interaction - gravitational waves, l'onde gravifique, as he called them (Poincare 1905, 1906).

A few years later in his lecture on "New concepts of matter" Poincare wrote about including Planck's discovery of the quantum nature of electromagnetic radiation in the framework of future physics. Poincare thus could be rightfully regarded as the founder of that approach to gravitation now called the relativistic quantum field of gravitational interaction. Naturally, the gravidynamics or field gravity (FG) theory should take its place in the list of the field theories of fundamental physical interactions.

The field approach to gravitation was partly developed by leading physicists in a number of studies, among them Birkhoff (1943, 1944), Moshinsky (1950), Thirring (1961), and Kalman (1961). Attempts for a quantum description of the field approach were made by Bronstein (1936), Fierz & Pauli (1939), Ivanenko & Sokolov (1947), Feynman (1963, 1971), Weinberg (1965), Zakharov (1965), and Ogievetsky \& Polubarinov (1965).

The strategy and basic principles of the field gravity theory were discussed by Feynman, who emphasized that "geometrical interpretation is not really necessary or essential for physics" (Feynman, Morinigo & Wagner 1995, p. 113). He pointed to the central role of the energy of the gravity field for a reasonable theory of gravitational interaction. Feynman's notorious words in a letter to his wife "Remind me not to come to any more gravity conferences" are related to this very issue, he did not wish to discuss the question of whether there is energy of the gravity field. For him gravitons were particles carrying the energy-momentum of the field: "the situation is exactly analogous to electrodynamics - and in the quantum interpretation, every radiated graviton carries away an amount of energy ℏω" (Feynman, Morinigo & Wagner 1995, p. 220).

A consistent field gravity theory, where the inertial frames, Minkowski space and localizable positive energy of the gravity field have the central role, has been partly developed by Sokolov and Baryshev (e.g.

Sokolov \& Baryshev 1980, Baryshev \& Sokolov 1983,1984; Sokolov 1992a-d; Baryshev 2003, 2006b) and will be presented below.

*Why FG is principally different from GR .*

An important note about the field approach should be done. The history of the field gravity is characterized by misleading claims and it demonstrates how hard it may be to create and develop scientific ideas. There are many papers and discussions about the derivation of Einstein's field equations from the spin 2 theory (another name for the field theory), and hence about the identity of general relativity and field approach. Feynman in his lectures on gravitation also tried to derive the full Einsteinian Lagrangian by iterating the Lagrangian of the spin 2 field, and later many studies have been made on this subject.

Misner, Thorne \& Wheeler (1973, chapter 7, p.178) wrote that "tensor theory in flat spacetime is internally inconsistent; when repaired, it becomes general relativity". They refer to papers by Feynman (1963), Weinberg (1965), and Deser (1970) on a "field" derivation of Einstein's equations.

However, Straumann (2000, p.16) pointed out the internal inconsistence of such attempts to derive Einstein's equations from the spin 2 field theory: 1) general relativity having black hole solutions violates the simple topological structure of the Minkowski space of the field gravity, and 2) general relativity has lost the energy-momentum tensor of the gravity field together with the conservation laws, which is the direct consequences of the global symmetry of the Minkowski space.

Padmanabhan (2004) gave a comprehensive review of all such attempts and demonstrated that all derivations of general relativity from a spin 2 field are based on some additional assumptions that are equivalent to the geometrization of the gravitational interaction. Indeed, general relativity and field gravity rest on incompatible physical principles, such as non-inertial frames and Riemann geometry of curved space on the one side, and inertial frames with Minkowski geometry of flat space on the other side.

Geometrical approach eliminates the gravity force, as already de Sitter (1916a) noted: "Gravitation is thus, properly speaking, not a 'force' in the new theory". This however leads to the problem of energy just because the work done by force changes the energy. Within the field approach the gravity force is directly defined in an ordinary sense as the fourth interaction, which has a quantum nature (Feynman 1971).

**4. Gravity physics in Poincare-Feynman's field approach**

In the Poincare-Feynman field approach, the gravity force between Newton's apple and the Earth is caused by the exchange of gravitons. Gravitons (real and virtual) are mediators of the gravitational interaction and represent quanta of a relativistic tensor field $\psi^{ik}$ in Minkowski space with metric $\eta^{ik}$.

The field approach offers a natural solution to the energy problem. Minkowski space implies the invariance under the Poincare group transformation and hence the usually defined energy-momentum tensor of the gravity field, as follows from Noether's theorem.

We stress that the construction of the field gravity (FG) is not yet completed and important questions are still open. For example, the quantization of the gravity field needs to take into account the conservation of the gravitational energy and the finiteness of the gravity force, in order to overcome the problem of non-renormalizability. The main strategy of the consistent field approach is not to write down the final non-linear exact equations, but to control the physical sense of all theoretical quantities used in the description of the gravitational interaction.

The consistent Lagrangian field gravity theory was started in the works by Thirring (1961) and Kalman (1961), and continued by Sokolov, Baryshev and others. Up to now, within the field gravity theory only the weak field approximation at the post-Newtonian level has been studied in detail, but this is enough to show the feasibility of this approach and to give predictions, which distinguish FG and GR.
Hence, the field gravity theory is not experimentally equivalent to the geometrical general relativity.
Below we follow the work by Baryshev (2003).

*Initial principles of the field gravity theory*

*The unity of the fundamental interactions.* Poincare (1905, 1906) suggested that all physical forces, including gravitation, could be considered within the same physical principles (especially the Lorentz invariance). The field approach actually continues this program. In Feynman's *Lectures on Gravitation* (Feynman, Moringo \& Wagner 1995) gravitation is described as a relativistic tensor field in Minkowski space, using the Lagrangian formalism of the field theory. Feynman discussed a standard quantum field description of gravity "just as the next physical interaction". He emphasized that "the geometrical interpretation is not really necessary or essential to physics".

There are several common physical elements that appear in the description of all fundamental physical interactions: the inertial reference frames; the flat Minkowski space-time; Lagrangian formalism; the existence of the energy-momentum tensor (EMT) of the field; the positive energy density of the field; the zero trace of the EMT for massless fields; the uncertainty principle; the superposition principle; the quanta of the energy of the field. These elements are the basis of the consistent field approach to gravity and form a natural starting point for understanding the physics of gravity phenomena similarly to other fundamental forces.

*The principle of consistent iterations.* The gravity field has a positive energy and this energy, in turn, becomes a new source of an additional gravity field and so on. This non-linearity is taken into account by an iteration procedure. It is usual in physics to consider first a linear approximation and then add non-linearity by means of iterations.

The field gravity theory is constructed step by step using an iteration procedure so that at each step all physical properties of the EMT of the gravity field are under control. Each step of iteration is described by linear gauge-invariant field equations with fixed sources in the right-hand side. An important outcome of this procedure is that the superposition principle can be reconciled with the non-linearity of the gravity field.

*The principle of least action.* The mathematical tool is the Lagrangian formalism of the relativistic field theory. To derive the equations of motion for the gravity field and for the matter one uses the principle of least action, which states that for the true motion the variation of the action $\delta S = 0$.

The action integral for the whole system of a gravitational field and particles (matter) within it, consists of three parts:

$$S = S_{(g)} + S_{(int)} + S_{(m)} = \frac{1}{c} \int (\Lambda_{(g)} + \Lambda_{(int)} + \Lambda_{(m)}) \, d\Omega \ . \tag{5}$$

The notations (g), (int), (m) refer to the actions for the free gravity field, the interaction, and the free particles (matter). The physical dimension of each part of the action (5) is [S]= [energy density] x [volume] x [time], meaning that the definition of energy is assumed within the principle of least action. Note that in general relativity the action integral has only two parts $S_{(g)}$ and $S_{(m)}$, and there is no Interaction part, because of the principle of geometrization.

*Lagrangian for free gravitational field.* Within the Poincare-Feynman approach the gravity field is presented by symmetric 2nd rank tensor potentials $\psi^{ik}$ with the trace $\psi = \eta_{ik} \psi^{ik}$ in Minkowski space with metric $\eta^{ik}$. The Lagrangian for a free gravitational field we take in the form:

$$\Lambda_{(g)} = -\frac{1}{16\pi G} \left[ \left( 2\psi_{nm}^{;n} \psi_{,l}^{lm} - \psi_{lm,n} \psi^{lm,n} \right) - \left( 2\psi_{ln}^{;l} \psi^{,n} - \psi_{,l} \psi^{,l} \right) \right] \ . \tag{6}$$

This differs from Thirring's (1961) choice by a divergent term, which does not change the field equations, but has the advantage that the canonical energy momentum tensor is symmetric, $\psi_{,l}^{ik} = \partial \psi^{ik} / \partial x^l$ is the ordinary partial derivative of the symmetric second rank tensor potential.

*Lagrangian for matter.* The Lagrangian for matter depends on the physical problem in question (particles, fields, fluid or gas). Gravity is also a kind of matter and at each iteration step it is considered as a source fixed by the preceding step. For relativistic point particles the Lagrangian is

$$\Lambda_{(p)} = -\eta_{ik} T_{(p)}^{ik} \ , \tag{7}$$

where $T_{(p)}^{ik}$ is the EMT of the particles

$$T_{(p)}^{ik} = \sum_a m_a c^2 \, \delta(\vec{r} - \vec{r}_a) \left\{ 1 - \frac{v_a^2}{c^2} \right\}^{1/2} u_a^i u_a^k \ . \tag{8}$$

Here *m, v, $u^i$* are the mass, 3-velocity, and 4-velocity of a particle. For a relativistic macroscopic body the EMT is

$$T^{ik}_{(m)} = (\varepsilon + p) u^i u^k - p \eta^{ik} \ , \tag{9}$$

where $\varepsilon$ and $p$ are the energy density and pressure of a comoving volume element.

*The principle of universality of gravitational interaction.* In the field approach the principle of universality states that the gravitational field $\psi^{ik}$ interacts with all kinds of matter via their energy-momentum tensor $T^{ik}$, so the Lagrangian for the interaction has the form:

$$\Lambda_{(int)} = -\frac{1}{c^2} \psi_{ik} T^{ik} \ . \tag{10}$$

The principle of universality, eq.(10), was introduced by Moshinsky (1950). It replaces the equivalence principle used in the geometrical approach. From the principle of universality of gravitational interaction (UGI) and the least action principle one can derive those consequences of the equivalence principle, which do not create paradoxes. According to UGI the free fall acceleration of a body does not depend on its rest mass.

The equivalence principle of GR cannot be a basis of the field gravity, because it eliminates the gravity force and accepts the equivalence between the inertial motion and the accelerated motion under gravity. E.g., the equivalence principle creates a puzzle for an electric charge resting in the gravity field on a laboratory table. Due to the equivalence of the laboratory frame (together with the table) and an accelerated frame with $a = g$, the charge on the table is equivalent with an accelerated charge and should radiate energy according to the relation $P = (2/3)(e^2/c^3) \, a^2$ ergs/s. In the field theory this charge does not radiate, because it is at rest in an inertial system. The concept of an inertial frame is preserved in field gravity and inertial and accelerated motions are not equivalent.

Instead of the GR principle of equivalence, the field gravity is based on the principle of universality of gravitational interaction, according to which gravity "see" only the energy momentum tensor of any matter. This point is also different from all "effective geometry" theories where the universality of gravity is understood as geodesic motion in Riemannian space.

### Basic equations of the field gravity theory

*Field equations.* Using the variation principle to obtain the field equations from the action (5) one must assume that the sources $T^{ik}$ of the field are fixed (or the motion of the matter is given) and vary only the potentials $\psi^{ik}$ (serving as the coordinates of the system). On the other hand, to find the equations of motion of the matter in the field, one should assume the field to be given and vary the trajectory of the particle (matter). So keeping the EMT of matter in (10) fixed and varying $\delta \psi^{ik}$ in (5) we get

$$-\psi^{ik,l}_{\ \ \ l} + \psi^{il,k}_{\ \ \ l} - \psi^{kl,i}_{\ \ \ l} - \psi^{,ik} - \eta^{ik}\psi^{lm}_{\ \ ,lm} + \eta^{ik}\psi^{,l}_{\ \ l} = \frac{8\pi G}{c^2} T^{ik} \ . \tag{11}$$

The field equations (11) are identical to the linear approximation of Einstein's field equations and that is why there are many similarities between GR and FG in the weak field regime.

However, the difference is that $\psi^{ik}$ and $\eta^{ik}$ both are true tensors and hence their sum $\psi^{ik} + \eta^{ik} = f^{ik}$ is the true tensor in Minkowski space. It means that for covariant components we have $\psi_{ik} + \eta_{ik} = f_{ik}$. But in GR for a weak field approximation the metric tensor presented as a sum $g^{ik} = \eta^{ik} + h^{ik}$, where quantities $h^{ik}$ and $\eta^{ik}$ are not tensors of a general Riemannian space. E.g. they change sign in covariant components due to the exact relation $g^{ik}g_{ik} = 4$, which is valid for any metric tensor, so that $g_{ik} = \eta_{ik} - h_{ik}$.

*Remarkable features of the field equations.* First, the divergence of the left side of the field equations (11) is zero, implying the conservation law

$$T^{ik}_{\ ,k} = 0 \ , \tag{12}$$

in the approximation corresponding to the considered step in the iteration procedure. In the zero approximation it does not include the EMT of the gravity field, but the first approximation contains the gravity field of the zero approximation and expresses the conservation laws and the equations of motion at the post-Newtonian level.

Second, the equations (11) are gauge invariant, i.e. they do not change under the following transformations of the potentials:

$$\psi^{ik} \Rightarrow \psi^{ik} + \lambda^{i,k} + \lambda^{k,i} \ . \tag{13}$$

An important difference between this gauge transformation and the general covariant transformation of coordinates in GR is that (13) performed in a fixed inertial reference frame.

Third, the gauge freedom (13) allows one to put four additional conditions on the potentials, in particular the Hilbert-Lorentz gauge:

$$\psi^{ik}_{,k} = \frac{1}{2} \psi^{,i} \ . \tag{14}$$

With the gauge (14) the field equations get the form of the wave equation:

$$\left( \Delta - \frac{1}{c^2} \frac{\partial^2}{\partial t^2} \right) \psi^{ik} = \frac{8\pi G}{c^2} \left[ T^{ik} - \frac{1}{2} \eta^{ik} T \right] \ . \tag{15}$$

The trace of this equation gives the wave equation for the scalar part $\psi = \eta_{ik} \psi^{ik}$ of the gravitational potentials:

$$\left( \Delta - \frac{1}{c^2} \frac{\partial^2}{\partial t^2} \right) \psi(\vec{r},t) = -\frac{8\pi G}{c^2} T(\vec{r},t) \ . \tag{16}$$

Note the opposite signs in the right-hand sides of eqs.(15, 14). We'll see that this corresponds to the important fact that the pure tensor part of the field represents attraction, while the scalar part gives repulsion. Another related thing is that in the Lagrangian (6) the tensor and scalar parts have opposite signs.

*The bi-component structure of the gravity field.* The multi-component structure of the tensor potential is a most important thing in the quantum field theory. It is well known that the symmetric 2nd rank tensor field $\psi^{ik}$ can be decomposed under the Poincare group into a direct sum of subspaces. It represents one particle with spin 2, one particle with spin 1, and two particles with spin 0 (Fronsdal 1958; Barnes 1965):

$$\{\psi^{ik}\} = \{2\} \oplus \{1\} \oplus \{0\} \oplus \{0'\} \ . \tag{17}$$

The tensor $\psi^{ik}$ contains n =10 independent components. The relation n = 2s + 1 between the number of components n and the value of the spin s is fulfilled for the four particles as 10 = 5+3+1+1 in eq.(17). As the field equations (11) are gauge invariant under the transformation (13) one may use 4 additional functions $\lambda^i$ to delete the 4 (from total 10) components corresponding to spin 1 and one of the spin 0, leaving only the spin 2 and the second spin 0 parts of the tensor potential. Hence, after the Hilbert-Lorentz gauge (14), the field equations (15) will describe the mixture of two fields with spin 2 and spin 0, generated by two corresponding parts of the source of the gravity field (Sokolov, Baryshev 1980):

$$\{\psi^{ik}\} = \{2\} \oplus \{0\} \quad \Leftrightarrow \quad \{T^{ik}\} = \{2\} \oplus \{0\} \ . \tag{18}$$

Now we can present the initial tensor potential and the EM tensor of the source as the sum of pure tensor spin 2 and pure scalar spin 0 parts:

$$\psi^{ik} = \psi^{ik}_{\{2\}} + \psi^{ik}_{\{0\}} = (\psi^{ik} - \frac{1}{4}\eta^{ik}\psi) + \frac{1}{4}\eta^{ik}\psi \quad \Leftrightarrow \quad T^{ik} = T^{ik}_{\{2\}} + T^{ik}_{\{0\}} = (T^{ik} - \frac{1}{4}\eta^{ik}T) + \frac{1}{4}\eta^{ik}T \ . \tag{19}$$

Hence, eq.(15) can be written in the form

$$\left( \Delta - \frac{1}{c^2} \frac{\partial^2}{\partial t^2} \right) \psi^{ik}_{\{2\}} = \frac{8\pi G}{c^2} T^{ik}_{\{2\}} \ , \quad \text{or} \quad \left( \Delta - \frac{1}{c^2} \frac{\partial^2}{\partial t^2} \right) \psi^{ik}_{\{2\}} = \frac{8\pi G}{c^2} \left[ T^{ik} - \frac{1}{2}\eta^{ik} T \right] \ . \tag{20}$$

and

$$\left(\Delta - \frac{1}{c^2}\frac{\partial^2}{\partial t^2}\right)\psi^{ik}_{\{0\}} = -\frac{8\pi G}{c^2}T^{ik}_{\{0\}}, \quad \text{or} \quad \left(\Delta - \frac{1}{c^2}\frac{\partial^2}{\partial t^2}\right)\psi\frac{1}{4}\eta^{ik} = -\frac{8\pi G}{c^2}T\frac{1}{4}\eta^{ik}. \quad (21)$$

This means that the field gravity theory is actually a tensor-scalar theory, where the scalar part of the field is simply the trace of the tensor potentials $\psi = \eta_{ik}\psi^{ik}$ generated by the trace of the energy-momentum tensor of the matter $T = \eta_{ik}T^{ik}$. According to the wave equations for spin 2 and spin 0 fields, both kinds of gravitons are massless particles.

Zakharov (1965) demonstrated that the tensor gravitational field in eq.(15) is described by spin 2 and spin 0 gravitons. From quantum field considerations (the condition of transversality of the gravitational vertex) he concluded that only spin 2 gravitons may be emitted, which corresponds to quadrupole gravitational waves. However, according to the wave equation (21), the trace $T(\vec{r}, t)$ of the EMT of matter will generate gravitational radiation, e.g. via spherical pulsations of a gravitating system. The radiated scalar gravitational wave is monopole and has a longitudinal character in the sense that a test particle in the wave moves along the direction of the wave propagation.

*Equations of motion for test particles.* Variation of the action integral (5) with respect to particle coordinates gives the equation of motion in a fixed gravitational field (Kalman 1961; Baryshev 1986):

$$A^i_k \frac{d(mcu^k)}{ds} = -mc B^i_{kl} u^k u^l, \quad (22)$$

where $mcu^k = p^k$ is the 4-momentum of the particle, and

$$A^i_k = \left(1 - \frac{1}{c^2}\psi_{ln}u^l u^n\right)\eta^i_k - \frac{2}{c^2}\psi_{kn}u^n u^i + \frac{2}{c^2}\psi^i_k,$$

$$B^i_{kl} = \frac{2}{c^2}\psi^i_{k,l} - \frac{1}{c^2}\psi_{kl}{}^{,i} - \frac{1}{c^2}\psi_{kl,n}u^n u^i.$$

The rest mass m of the particle appears in both sides, so it cancels off. This shows how from the principles of least action and universality follows the principle of equivalence in the form: the rest mass m of a body is equal to its inertial and gravitational masses $m = m_I = m_G$. The rest mass includes all contributions from all interactions. Hence a test of the equivalence principle, when masses consist of different chemical materials, in fact checks the universality of all contributions into the rest mass.

*Repulsive force of the scalar part.* Inserting to the equation of motion (22) the scalar part of the gravitational potentials $\psi^{lm}_{\{0\}} = (1/4)\eta^{lm}\psi$, we get the equations of motion of a particle in the scalar field as

$$\left(1 + \frac{1}{4c^2}\psi\right)\frac{dp^i}{ds} = \frac{m}{4c}\left(\psi^{,i} - \psi_{,l}u^l u^i\right), \quad (23)$$

In the case of a weak field $\psi/c^2 \ll 1$ this equation gives for spatial components (i=ά) the expression for the gravity force

$$\frac{d\vec{p}}{dt} = -\frac{m}{4}\vec{\nabla}\psi = \frac{1}{2}m\vec{\nabla}\varphi_N, \quad (24)$$

where we take into account that the trace of the weak static field is equal to $\psi = -2\varphi_N$.

This means that the scalar (spin 0) part of the tensor field leads to a repulsive force and only together with the pure tensor (spin 2) part the result is the Newtonian force

$$\vec{F}_N = \vec{F}_{\{2\}} + \vec{F}_{\{0\}}. \quad (25)$$

## 5. The energy-momentum tensor of the gravity field

The standard Lagrangian formalism of the relativistic field theory gives for the Lagrangian of the gravity field (6) the following expression for the canonical energy-momentum tensor:

$$T^{ik}_{(g)} = \frac{1}{8\pi G}\left[\left(\psi^{lm,i}\psi_{lm}{}^{,k} - \frac{1}{2}\eta^{ik}\psi_{lm,n}\psi^{lm,n}\right) - \frac{1}{2}\left(\psi^{,i}\psi^{,k} - \frac{1}{2}\eta^{ik}\psi_{,l}\psi^{,l}\right)\right] . \qquad (26)$$

Two important remarks should be made about this expression. First, the EMT has an ordinary tensor character and so it is conceptually well defined. However, the Lagrangian formalism cannot give a unique expression for an EMT of any field (e.g. Bogolyubov \& Shirkov 1976; Landau \& Lifshitz 1971) because a term with zero divergence can always be added. For the final determination of the EMT of the field additional physical requirements must be used, like the positive energy density, the symmetry, and zero value for the trace in the case of a massless field.

Second, the negative sign of the scalar part (the 2nd term in brackets) does not mean the spin 0 field has negative energy. It reflects the repulsive force produced by the scalar when the field interacts with the sources, so that the energy of scalar field effectively compensate the energy of the tensor field.

For the free field the energy is positive for both the pure tensor (spin 2) and scalar (spin 0) components. Indeed, the total Lagrangian (6) for the case of a free field can be divided into two independent parts that correspond to two independent particles with spin 2 ($\phi^{ik}$) and spin 0 ($\psi$). This gives the following free field Lagrangians (Sokolov, Baryshev 1980):

$$\Lambda_{\{2\}} = \frac{1}{16\pi G}\phi_{ik,l}\phi^{ik,l} \qquad \text{and} \qquad \Lambda_{\{0\}} = \frac{1}{64\pi G}\psi_{,l}\psi^{,l} . \qquad (27)$$

Both signs are positive due to the positive energy density condition for integer spin free particles. Corresponding EMTs for the tensor and scalar free fields are

$$T^{ik}_{\{2\}} = \frac{1}{8\pi G}\phi_{lm}{}^{,i}\phi^{lm,k} \qquad \text{and} \qquad T^{ik}_{\{0\}} = \frac{1}{32\pi G}\phi^{,i}\phi^{,k} . \qquad (28)$$

These EMTs are symmetric, have a positive energy density and a zero trace for the case of plane monochromatic waves.

*The role of the scalar part of the field.* The scalar $\psi$ is an intrinsic part (the trace) of the gravitational tensor potential $\psi^{ik}$ and this is radically different from an extra scalar field $\varphi$, which is introduced in the Jordan-Brans-Dicke theory. So the observational constraints existing for this extra scalar field do not restrict the scalar part $\psi$ of the tensor field $\psi^{ik}$. Moreover, without the scalar $\psi$ it is impossible to explain the classical relativistic gravity effects (this is considered in the next paper).

The most intriguing consequence of the field gravity theory is that the scalar part (spin 0) corresponds to a repulsive force, while the pure tensor part (spin 2) corresponds to attraction. This explains the "wrong" sign for the scalar part in the Lagrangian for the gravity field (6).

## 6. Conclusion: detection of gravity energy as a crucial test of gravity physics

We have demonstrated that the physical understanding of the gravitational interaction in geometrical and field approaches is dramatically different. The consistent field approach predicts that the gravity force has an ordinary quantum nature and actually is presented by the sum of the attraction (spin 2) and repulsion (spin 0) components, which opens new possibilities for experimental study of the gravity. For practical cosmology, observational or experimental tests capable of distinguishing between these alternative ways to understand gravitational interaction are important, because the implications for cosmological models are dramatic.

In particular, the crucial test of gravity physics will be a detection of the energy of gravity field both in static gravitating systems and in the case of free field in the form of gravitational waves. In the next paper we show that the classical relativistic gravity effects in the weak field are identical in both theories, though there are also specific effects of the field gravity which may distinguish it from general relativity: scalar gravitational waves, the translational motion of rotating bodies, the surface and the magnetic field of the relativistic compact bodies in "black hole candidates", and the cosmological gravitational redshift.